\documentclass[reqno]{amsart}

\usepackage[style=alphabetic,sorting=none,backend=biber,maxbibnames=99]{biblatex}
\bibliography{guest0x0}
\usepackage{tiz-note}
\usepackage[hyphenbreaks]{breakurl}

\begin{document}

\title{A tutorial on implementing De Morgan cubical type theory}

\author{Tesla Zhang}
\address{Carnegie Mellon University}
\email{teslaz@cmu.edu}
\urladdr{https://personal.psu.edu/yqz5714}
\date\today

\newcommand{\GuestName}{\textsc{Guest0x0}}
\newcommand{\CTT}{{\mancube}TT}
\newcommand{\II}{\mathbb{I}}
\newcommand{\concat}{\textsf{concat}}
\newcommand{\sym}{\textsf{sym}}
\newcommand{\PGvdash}{\Psi;\Gamma\vdash}
\newcommand{\PiGvdash}[1]{\Psi,#1:\II;\Gvdash}
\newcommand{\PGfvdash}[1]{\Psi;\Gamma,#1\vdash}
\newcommand{\lcon}{\textsf{0}}
\newcommand{\rcon}{\textsf{1}}
\newcommand{\cond}{\textsf{cond}}
\newcommand{\face}{\textsf{face}}
\newcommand{\Partial}[2]{\textsf{Partial}_{#1}~{#2}}
\newcommand{\conj}{\theta}
\newcommand{\disj}{\vartheta}
\newcommand{\PathTy}[3]{\textsf{Path}_{#1}~{#2}~{#3}}
\newcommand{\ExtTy}[4]{\textsf{Ext}_{#1}~{#2}~\LRbbar{\begin{array}{#3}#4\end{array}}}
\newcommand{\extTy}[3]{\textsf{Ext}_{#1}~{#2}~\lrbbar{#3}}
\newcommand{\papp}[2]{{#1}~@~{#2}}
\newcommand{\SubTy}[4]{\textsf{Sub}_{#2}~{#1}~\LRbbar{\begin{array}{#3}#4\end{array}}}
\newcommand{\subTyImpl}[3]{\textsf{Sub}_{#2}~{#1}~{#3}}
\newcommand{\subTy}[3]{\subTyImpl{#1}{#2}{\lrbbar{#3}}}
\newcommand{\inS}[2]{\textsf{inS}_{#1}(#2)}
\newcommand{\outSLong}{\textsf{outS}}
\newcommand{\outS}[2]{\outSLong_{#1}(#2)}
\newcommand{\isCond}[1]{\fbox{\(#1\)}~\fs{valid}}
\newcommand{\isSatisfied}[1]{#1~\fs{satisfied}}
\newcommand{\coeLong}{\textsf{coerce}}
\newcommand{\coe}[2]{\textsf{coe}_{#1}^{#2}}
\newcommand{\hcompLong}{\textsf{hcomp}}
\newcommand{\hcomp}[3]{\hcompLong_{#1}{(#2, #3)}}

% carlo: https://gist.github.com/cangiuli/b21c9f8cb49dde06eef6480c29f7cf21
\pgfdeclarelayer{frontmost}
\pgfsetlayers{main,frontmost}
\usetikzlibrary{patterns}

\tikzset{
  carlo-axes/.style =
  {
    y = {(0,-1)},
    z = {(-0.6,0.6)}
  } ,
  shorten <>/.style =
  {
    shorten >=#1 , shorten <=#1
  } ,
  equals arrow/.style =
  {
    arrows = - ,
    double equal sign distance ,
  } ,
}

\newcommand{\carloCubeBullets}{
\begin{pgfonlayer}{frontmost}\begin{scope}[every node/.style = {inner sep = 0pt}]
\foreach \x in {0,1}
 \foreach \y in {0,1}
  \foreach \z in {0,1}
   \node (\x\y\z) at (\x , \y , \z) {\textbullet} ;
\end{scope}\end{pgfonlayer}}
\newcommand{\carloSqVertices}[1]{
\begin{pgfonlayer}{frontmost}\begin{scope}[every node/.style = {inner sep = 0pt}]
\foreach \x in {0,1}
 \foreach \y in {0,1}
  \node (\x\y) at (\x , \y) {#1} ;
\end{scope}\end{pgfonlayer}}
\newcommand{\carloSqBullets}{\carloSqVertices{\textbullet}}
\newcommand{\carloTikZ}[1]{
  \begin{tikzpicture}[carlo-axes, scale = 1.6]
		#1
	\end{tikzpicture}}
\newcommand{\carloCoord}[3]{\begin{scope}[shift={(-1.5,0)},scale=0.4]
    \draw [->] (0,0,0) to node [above,pos=1] {#1} (1,0,0) ;
    \draw [->] (0,0,0) to node [right,pos=1] {#2} (0,1,0) ;
    \draw [->] (0,0,0) to node [above,pos=1.3] {#3} (0,0,1) ;
  \end{scope}}
\newcommand{\carloCoordSq}[2]{\begin{scope}[shift={(-1.5,0)},scale=0.4]
    \draw [->] (0,0) to node [above,pos=1] {#1} (1,0) ;
    \draw [->] (0,0) to node [right,pos=1] {#2} (0,1) ;
  \end{scope}}
\newcommand{\carloXyz}{\carloCoord{$x$}{$y$}{$z$}}
\newcommand{\carloXy}{\carloCoordSq{$x$}{$y$}}
\newcommand{\refcube}[1]{\ExecuteMetaData[cubes]{#1}}
\newcommand{\carloCTikZ}[1]{\begin{center}
\carloTikZ{#1}
\end{center}}
\newcommand{\shiftTikZ}[2]{\begin{scope}[shift={(#1)}]
#2
\end{scope}}

% Comparison
\newcommand{\diffA}[1]{{\color{constant} #1}}
\newcommand{\diffB}[1]{{\color{comment} #1}}

\begin{abstract}
This tutorial explains (one way) how to implement De Morgan cubical type theory
to people who know how to implement a dependent type theory. It contains an introduction
to basic concepts of cubes, type checking algorithms under a cofibration,
the idea of ``transportation rules'' and cubical operations.
This tutorial is a by-product of an experimental implementation\footnotemark
% https://tex.stackexchange.com/a/67590/145304#comment784980_67590
of cubical type theory,
called \GuestName.
\end{abstract}
\maketitle
\stepcounter{footnote}\footnotetext{\url{https://github.com/ice1000/guest0x0}}
\tableofcontents
\section{Introduction}
Throughout this tutorial, \fbox{boxes} will be used in the following two cases:
\begin{itemize}
\item to clarify the precedences of symbols when formulae become too large. \\
e.g. \fbox{$\Gvdash \fbox{$\lam x M$}~:~\fbox{$(y:A)\to B$}
\Leftarrow \fbox{$\lam x u$}$}.
\item to distinguish type theory terms from natural language text. \\
e.g. ``combine a term \fbox{$a$} with a term \fbox{$b$} to get a term \fbox{$a~b$}''.
\end{itemize}
\subsection{Target Audience}
This tutorial assumes familiarity with the following:
\begin{itemize}
\item Dependent type theory concepts, such as formation rules,
introduction rules, eliminators, etc., and functional programming.
\item Programming and theorem proving in a proof assistant based on dependent type theories.
\item The ability to translate (simple) typing rules into a type-checking procedures
and combine them into an algorithm.
\item Basic understanding of De Morgan cubical type theory~\cite{CCHM,CHM}
(hereafter as \CTT{}), including the interval type $\II$, the path type,
the idea of representing $n$-dimensional cubes using terms with interval variables in it,
and the De Morgan operators on $\II$.
\end{itemize}
This tutorial will not treat substitution formally --
variable names are assumed to respect capture-avoiding substitution.
In the implementations, any binding representation that works for untyped
$\lambda$-calculus should work for the type theory introduced in this tutorial.
\begin{notation}\label{not:pre}
We will prefer using $x, y, z$ for variables
and other Latin letters like $u,v,a,b,c,A,B,C$ for terms
(preferably uppercase for types and lowercase for terms).

Instead of the more traditional \fbox{$\lambda x.b$},
the notation for $\lambda$-abstraction is \fbox{$\lam x b$} 
following the style of the Arend language.
We will also use the conventional shorthand
\fbox{$(x:A)~(y:B)\to C$} for nested $\Pi$-types.

Substitution is denoted by $u[v/x]$.
One may think of this notation as ``fractional multiplication'' $u\times \frac v x$,
where the denominator $x$ is cancelled out from $u$ and \textit{replaced} with the numerator $v$.
Other authors may use $u[x\mapsto v]$, $u[x:=v]$, $[v/x]u$, etc.

Definitional equality (a.k.a. judgmental equality) is denoted $u\equiv v$.
\end{notation}
\begin{notation}\label{not:syntax-def}
We will write \fbox{$u, A::=$} for syntax definition of terms,
and define the syntax of \CTT{} by extending the syntax of Martin-L\"of type theory
with a few term former at a time, instead of putting everything together
in a single, unified BNF grammar.
The typing rules will be introduced similarly.

We will extend the BNF grammar with the \textit{list} operator like \fbox{$\overline{(x_i:A_i)}\to B$},
which means that the string below the line can be repeated one or more times,
optionally indexed by a subscript such as $i$.
\end{notation}
Some quick warm-ups:
\begin{exercise}
Translate the following typing rules into an algorithmic description:
\begin{mathpar}
\inferrule{\Gamma,x:A\vdash b:B[x/y]}{\Gvdash \fbox{$\lam x b$} : (y:A)\to B} \and
\inferrule{\Gvdash u : (x:A)\to B \\ \Gvdash v:A}{\Gvdash u~v:B[v/x]}
\end{mathpar}
Which of these is an introduction rule, and which is an elimination rule?
\end{exercise}
\begin{exercise}\label{ex:concat-sym}
Consider path concatenation and symmetry in \CTT{}:
\begin{align*}
\concat&:(p:a=b)\to(q:b=c)\to{a=c}\\
\sym&:(p:a=b)\to{b=a}
\end{align*}
Define both using \hcompLong{} on the following squares,
preferably in a cubical programming language:
% https://q.uiver.app/?q=WzAsOCxbMCwwLCJhIl0sWzEsMCwiYyJdLFswLDEsImEiXSxbMSwxLCJiIl0sWzIsMSwiYSJdLFsyLDAsImIiXSxbMywxLCJhIl0sWzMsMCwiYSJdLFswLDEsIlxcY29uY2F0KHAsIHEpIiwwLHsic3R5bGUiOnsiYm9keSI6eyJuYW1lIjoiZGFzaGVkIn19fV0sWzIsMCwiIiwwLHsibGV2ZWwiOjIsInN0eWxlIjp7ImhlYWQiOnsibmFtZSI6Im5vbmUifX19XSxbMiwzLCJwIiwyXSxbMywxLCJxIiwyXSxbNCw1LCJwIl0sWzQsNiwiIiwyLHsibGV2ZWwiOjIsInN0eWxlIjp7ImhlYWQiOnsibmFtZSI6Im5vbmUifX19XSxbNiw3LCIiLDIseyJsZXZlbCI6Miwic3R5bGUiOnsiaGVhZCI6eyJuYW1lIjoibm9uZSJ9fX1dLFs1LDcsIlxcc3ltKHApIiwwLHsic3R5bGUiOnsiYm9keSI6eyJuYW1lIjoiZGFzaGVkIn19fV1d
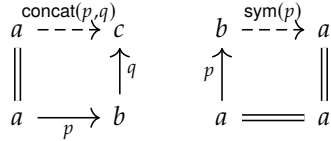
\[\begin{tikzcd}
	a & c & b & a \\
	a & b & a & a
	\arrow["{\concat(p, q)}", dashed, from=1-1, to=1-2]
	\arrow[Rightarrow, no head, from=2-1, to=1-1]
	\arrow["p"', from=2-1, to=2-2]
	\arrow["q"', from=2-2, to=1-2]
	\arrow["p", from=2-3, to=1-3]
	\arrow[Rightarrow, no head, from=2-3, to=2-4]
	\arrow[Rightarrow, no head, from=2-4, to=1-4]
	\arrow["{\sym(p)}", dashed, from=1-3, to=1-4]
\end{tikzcd}\]
\end{exercise}
\subsection{Motivation}
This tutorial is intended to help readers to get more familiar with how
\CTT{} works under the hood, what difficulties it is having in implementations,
what it can already do, and what it cannot do yet.

\CTT{} is a type theory evolved from a model using Kan cubical sets~\cite{CubicalSets},
which uses sophisticated homotopy theory. Computer scientists, on the other hand,
usually do not have relevant courses taught in their undergraduate program.
However, it is also the computer scientists who are supposed to implement \CTT{} as
programming languages. This tutorial tries to help those who did not study homotopy
theory, but have learned about the informal concepts of \CTT{}
(like what is written in~\cite{NCTT}) and wish to learn the implementation details of \CTT.

\CTT{} extends Martin-L\"{o}f type theory with a huge amount of new constructions,
especially the typing rules are written in a very compact way (like in~\cite{HCompPDF}).
This tutorial aims to discuss them from an algorithmic perspective,
and hopefully to inspire more people to implement cubical type theory,
to fuse these ideas into other work, or just to worship these brilliant ideas.

This tutorial is a by-product of an experiment in implementing \CTT, called \GuestName,
a project created to encourage a particular person to learn \CTT.
The story ends up in the worst way: the person did not learn \CTT, and instead created
a new project to encourage the author of \GuestName{} to learn extensional type theory.

\section{Type checking cubes}\label{sec:tyck-cube}
This section introduces the notion of \textit{partial elements} and
motivates typing judgments with cofibrations in the context.

\subsection{The interval $\II$ and contexts}\label{sub:interval}
\CTT{} has the interval type:
\[\vdash \isType\II\quad \vdash \lcon:\II \quad \vdash \rcon:\II\]
The interval type and its products are used to represent dimensions
(ignoring the De Morgan structures for now).
\begin{example}\label{ex:interval-in-ctx}
Suppose \fbox{$x:\II\vdash\isType{A}$} and \fbox{$x:\II\vdash u:A$}.
From a semantical or a topological perspective, one can say:
\begin{itemize}
\item $A$ is a (type) line between $A[\lcon/x]$ and $A[\rcon/x]$.
\item $u$ is a (term) line between $u[\lcon/x]$ and $u[\rcon/x]$.
\item The type of a line is a line, so the type of $u$ is $A$.
\end{itemize}
Then, because \emph{typing relations are preserved by substitution},
the following typing relations hold:
\begin{mathpar}
\inferrule{}{u[\rcon/x]:A[\rcon/x]}\and
\inferrule{}{u[\lcon/x]:A[\lcon/x]}
\end{mathpar}
We may visualize the fact as:
% https://q.uiver.app/?q=WzAsNixbMCwwLCJ1W1xcbGNvbi9pXSJdLFswLDEsInVbXFxyY29uL2ldIl0sWzIsMCwiQVtcXGxjb24vaV0iXSxbMiwxLCJBW1xcbGNvbi9pXSJdLFsxLDBdLFsxLDFdLFswLDEsInUiXSxbMiwzLCJBIl0sWzQsNSwiOiIsMSx7InN0eWxlIjp7ImJvZHkiOnsibmFtZSI6Im5vbmUifSwiaGVhZCI6eyJuYW1lIjoibm9uZSJ9fX1dXQ==
\[\begin{tikzcd}
	{u[\lcon/x]} & {} & {A[\lcon/x]} \\
	{u[\rcon/x]} & {} & {A[\rcon/x]}
	\arrow["u", from=1-1, to=2-1]
	\arrow["A", from=1-3, to=2-3]
	\arrow["{:}"{description}, draw=none, from=1-2, to=2-2]
\end{tikzcd}\]
\end{example}
From~\cref{ex:interval-in-ctx} we motivate the following notational
convention for contexts in \CTT, as in~\cref{not:ctx}.
\begin{notation}\label{not:ctx}
Typing judgments are written as \fbox{$\PGvdash \isType{A}$} and \fbox{$\PGvdash u:A$},
where \fbox{$\Psi;\Gamma$} is the usual \textit{context} in type theories,
with variables classified into two groups: if a variable has type $\II$,
it goes to $\Psi$, otherwise it goes to $\Gamma$.
This convention is borrowed from~\cite{ABCFHL}.
\end{notation}
Note that~\cref{not:ctx} does not imply that contexts has to be classified
in the implementations. The \GuestName{} type checker mix intervals
and other bindings in a unified context, just like usual dependent type checkers.
\begin{remark}\label{rem:line-ori}
Consider \fbox{$x:\II\vdash u:A$} and \fbox{$\vdash v:\II\to A$}.
Usually both are referred to as a \textit{line}, but they are very different.
Suppose the context is weakened with $y:\II$ to be a $2$-dimensional space,
in which $u$ exists as a line:
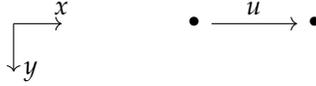
\carloCTikZ{\carloXy
\node (0) at (0 , 0) {\textbullet} ;
\node (1) at (1 , 0) {\textbullet} ;
\draw[->] (0) -- (1) node [midway, above] {$u$};}
Note that the orientation of $u$ is fixed to be horizontal.
However, for $v$, we can apply either $x$ or $y$ to get a line oriented differently:
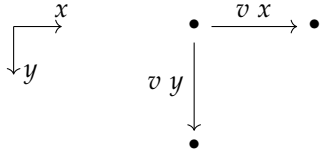
\carloCTikZ{\carloXy
\node (0) at (0 , 0) {\textbullet} ;
\node (1) at (1 , 0) {\textbullet} ;
\node (2) at (0 , 1) {\textbullet} ;
\draw[->] (0) -- (1) node [midway, above] {$v~x$};
\draw[->] (0) -- (2) node [midway, left] {$v~y$};}
So, interval application may also be thought of as
\textit{placing an $n$-dimensional cube at the given orientation}.
\end{remark}
\begin{remark}\label{rem:sq-ori}
Unlike lines as discussed in~\cref{rem:line-ori}, squares are much more flexible.
Consider \fbox{$\vdash u:\II\to\II\to A$} in a $2$-dimensional context,
there are already two different ways to place it:
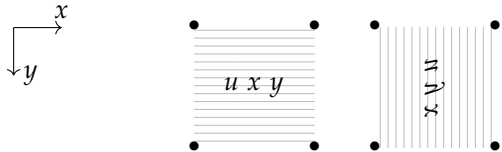
\carloCTikZ{\carloXy
\carloSqBullets
\fill [pattern color=lightgray,pattern=horizontal lines] (0,0) rectangle (1,1) ;
\node (c) at (0.5, 0.5) {$u~x~y$} ;
\shiftTikZ{1.5,0}{
\carloSqBullets
\fill [pattern color=lightgray,pattern=vertical lines] (0,0) rectangle (1,1) ;
\node[rotate=90,xscale=-1] (c) at (0.5, 0.5) {$u~y~x$} ;
}}
Note that these two placements are symmetric with respect to the diagonal.

In case of contexts and cubes of higher dimensions,
the situations are much more complicated.
For example, with one more dimension $z:\II$,
$u$ may be placed in three orientations:
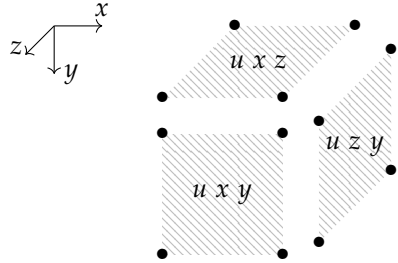
\carloCTikZ{\carloXyz
\refcube{TopUXY}
\shiftTikZ{0,0.3}{\refcube{FrontUXY}}
\shiftTikZ{1.3,0.2}{\refcube{LeftUXY}}
}
Note that all of them can also be reflected by their diagonals.
\end{remark}

\begin{defn}\label{defn:homo-path}
A term-line is said to be \textit{homogeneous} if its type is a constant type-line.
The notion of homogeneous term-line can be generalized to any $n$-cube of terms.
\end{defn}

\subsection{Partial elements}\label{sub:partial}
In \CTT{}, the idea that \textit{open shapes can be filled}
is the core concept that makes terms in type theory space-like,
and to do that, a \textit{composition operation} is added to \CTT{} as a structure.

The motivation is that with only the interval type and Martin-L\"of type theory,
we may not be able to describe every cube that makes geometric or topological sense.
For example, the squares in~\cref{ex:concat-sym} make perfect sense in
geometry or topology, but in \CTT{}, they have to be constructed using the composition operation.
The composition operation takes a description of some parts of an $n$-dimensional cube
(an \textit{incomplete} cube, or a \textit{partial} cube) and completes it.
To describe the input of composition, we introduce partial elements.
\begin{terminology}
In~\cite{CCHM}, partial elements are called \textit{systems},
but the sequel~\cite{CHM} switched to the name ``partial element''.
Cubical Agda~\cite{CubicalAgda} also uses ``partial element''.
\end{terminology}
When we write \fbox{$x:\II,y:\II\vdash u:A$},
we are describing the following 2-dimensional cube, which is a square:
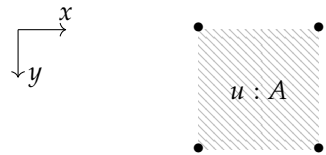
\carloCTikZ{\carloXy
\carloSqBullets
\fill [pattern color=lightgray,pattern=north west lines] (0,0) rectangle (1,1) ;
\node (center) at (0.5, 0.5) {$u:A$} ;}
For simplicity we assume \fbox{$\vdash \isType{A}$}, say, $A$ does not depend on $x$ or $y$.

Suppose we want to use the composition operation to create
a square \fbox{$x:\II,y:\II\vdash u:A$}
such that its top-left corner is a point $a:A$, and the line on its
right-hand side is a line $y:\II \vdash v:A$:
\carloCTikZ{\carloXy
\foreach \y in {0,1} \node (1\y) at (1 , \y) {\textbullet} ;
\node (00) at (0, 0) {$a$};
\node (01) at (0, 1) {\textbullet};
\draw (10) -- node [right] {$v$} (11); %
% \fill [pattern color=lightgray,pattern=north west lines] (0,0) rectangle (1,1) ;
% \node (center) at (0.5, 0.5) {$u:A$} ;
}

Translating that into type theory,
the goal is to construct a term $u$ such that the following holds:
\begin{align*}
u[\lcon/x, \lcon/y]&\equiv a\\
u[\rcon/x]&\equiv v
\end{align*}
The construction will be discussed in later sections,
and for now we focus on how to describe these partial boundaries
(also known as \textit{configurations} of a cube).
We introduce a straightforward syntax called \textit{partial elements}
for these cubes, e.g. the above partial element is written as:
\[\LRbbar{\begin{array}{rc}
  x=\lcon \land y=\lcon&\mapsto a\\
  x=\rcon &\mapsto v
\end{array}}\]
To define this formally, we need to define the syntax of the left-hand-side of $\mapsto$.
They are called \textit{cofibrations} in \CTT.
\begin{terminology}
In~\cite{CCHM}, cofibrations are called \textit{face restrictions},
which is more (geometrically) intuitive but also longer.
\end{terminology}
The syntax of a cofibration is defined to be a disjunction normal form
(a disjunction list of conjunctions) of face \textit{conditions}
like \fbox{$x=\lcon$} or \fbox{$x=\rcon$}, as in~\cref{fig:cofib} (recall~\cref{not:syntax-def}).
\begin{figure}[h!]
\[\begin{array}{rll}
  \cond ::= & x=\lcon \mid x=\rcon & \text{condition} \\
  \conj ::= & \cond~\overline{\land~\cond} & \text{conjunction} \\
  \disj ::= & \bot \mid \top \mid \conj~\overline{\lor~\conj} & \text{disjunction}
\end{array}\]
\caption{Syntax of cofibrations}\label{fig:cofib}
\end{figure}

The typing rules for cofibrations is also straightforward, as in~\cref{fig:tyck-cofib}.
\begin{figure}[h!]
\begin{mathpar}
\inferrule{(x:\II) \in \Psi}{\Psi\vdash \isCond{x=\lcon}} \and
\inferrule{(x:\II) \in \Psi}{\Psi\vdash \isCond{x=\rcon}}\and
\inferrule{\forall i. (\Psi\vdash \isCond{\cond_i})}{\Psi\vdash \bigwedge\nolimits_{i} \cond_i}\and
\inferrule{\forall i. (\Psi\vdash \conj_i)}{\Psi\vdash \bigvee\nolimits_{i} \conj_i}\and
\inferrule{}{\Psi\vdash \top} \and
\inferrule{}{\Psi\vdash \bot}
\end{mathpar}
\caption{Typing rules of cofibrations}\label{fig:tyck-cofib}
\end{figure}

The meaning of cofibrations is simple. Suppose we are in a $3$-dimensional context,
which means there are $3$ intervals in the context, i.e. \fbox{$\Psi:=x:\II,y:\II,z:\II$}.
Then:
\begin{itemize}
\item As mentioned before, the term \fbox{$\Psi\vdash u:A$} corresponds to
(the \textit{filling} of) a $3$-dimensional cube:
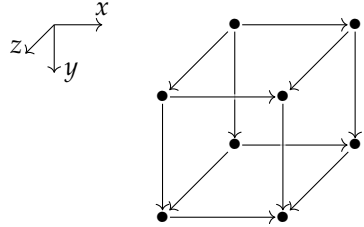
\carloCTikZ{\carloXyz \refcube{Empty}}
\item A $\cond$ specifies a $2$-dimensional cube (a square) face in $\Psi$, e.g. \fbox{$x=\lcon$}
corresponds to the following square:
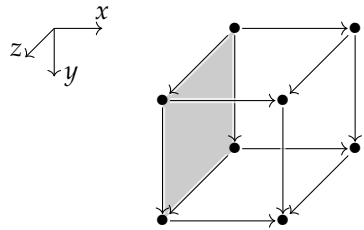
\carloCTikZ{\carloXyz \refcube{XEquivL}}
\item A $\conj$ specifies any $n$-cube (for $n\leq 2$) in $\Psi$, e.g.
\fbox{$x=\lcon\land y=\rcon$} specifies a line $v$,
and \fbox{$x=\lcon\land y=\rcon\land z=\lcon$} specifies a point $\star$:
\carloCTikZ{\carloXyz \refcube{XYLR}}
\item A $\disj$ talks about several $n$-cubes (for $n\leq 2$) in $\Psi$ at the same time, e.g.
\fbox{$x=\lcon\lor x=\rcon$} corresponds to the following two squares:
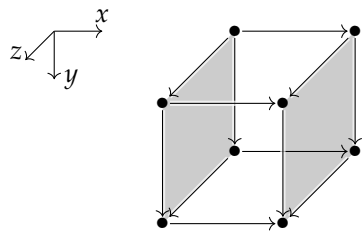
\carloCTikZ{\carloXyz \refcube{LRFaces}}
\item A $\bot$ cofibration is called the \textit{absurd} cofibration,
which specifies nothing.
\item A $\top$ cofibration is called the \textit{truth} cofibration,
which specifies everything.
\end{itemize}
Using cofibrations, we define the syntax of partial elements
by extending the syntax for terms in~\cref{fig:parEl}.
We define a special kind of partial elements $\lrbbar u$,
which is a partial element with a single face under the truth cofibration,
called \textit{trivial} partial elements.
These partial elements are actually not ``partial'', but ``full''
in the sense that \textit{every} face is assigned with a single term.
\begin{figure}[h!]
\[\begin{array}{rl}
    \face::=&\overline{\conj\mapsto u}\\
    u,A::=&\lrbbar{\face} \mid \lrbbar{u} \mid \cdots~\text{(other term formers, recall~\cref{not:syntax-def})}
\end{array}\]
\caption{Syntax of partial elements}\label{fig:parEl}
\end{figure}

Note that we also need to define the type of partial elements
(hereafter as \textit{partial types}),
and the type needs to contain the following two pieces of information:
\begin{itemize}
\item The faces being specified, $\disj$.
\item The type of the faces, $A$.
\end{itemize}
Thus we directly define the syntax of partial types as in~\cref{fig:parTy}:
\begin{figure}[h!]
\[u,A::= \Partial{\disj}{A} \mid \cdots~\text{(see~\cref{fig:parEl})}\]
\caption{Syntax of partial types}\label{fig:parTy}
\end{figure}

There are several advantages to arrange the cofibrations
as disjunction-normal forms:
\begin{itemize}
\item We put $\conj$ to the left-hand-side of $\mapsto$,
so every clause in a partial element specifies a single face.
\item It is easy to get the type of a partial element:
since each clause specifies a face using $\conj$,
their disjunction is the $\disj$ in the corresponding partial type.
\end{itemize}
It remains to derive the evaluation and typing rules for partial elements.

\subsection{Reducing partial elements by cofibrations}\label{sub:red-cofib}
We reduce well-typed (we will define well-typedness of
partial elements later in~\cref{sub:tyck-cofib}) partial elements by
iterating their face clauses. Consider the following face clause
with no conjunction:
\[x=\lcon\mapsto u\]
We may substitute the variable $x$ with three possible terms of type $\II$:
\begin{itemize}
\item Another variable $y$. In this case, we simply replace $x$ with $y$
and the face becomes $y=\lcon\mapsto u$.
\item $\lcon$, i.e. take the face that $x=\lcon$ in this partial element.
Then, evidently, we get the face $u$, so the partial element should reduce
to a trivial one $\lrbbar u$, with other faces ignored.
In this case, we say that the face is \textit{satisfied}.
\item $\rcon$, i.e. take the face that $x=\rcon$,
which is unspecified by this face, so we drop this particular face and
proceed with the rest of the faces.
In this case, we say that the face is \textit{contradicted}.
\end{itemize}
With the presence of conjunctions, we iterate through each $\cond$,
drop the conditions that are satisfied, and drop the faces
if one of their conditions is contradicted.

There are two more special cases in a conjunction $\conj$:
\begin{itemize}
\item It can be self-contradictory.
This happens when for a variable $x$, we have both $x=\lcon\in\conj$
and $x=\rcon\in\conj$. In this case, we also remove the face.
\item It can contain duplicated information.
For example, we may write $x=\lcon\land x=\lcon$.
It is encouraged to deduplicate these conditions for spatial efficiency.
In this tutorial, we assume deduplication of cofibrations everywhere.
\end{itemize}
By that we may claim the following:
\begin{prop}\label{prop:unique}
In a $\conj$ cofibration, every variable appears uniquely in a $\cond$.
\end{prop}

Substitution may also change partial types \fbox{$\Partial\disj A$}.
The rules are essentially the same as partial elements,
but in case of a face is satisfied, we reduce
the partial type into the annotated type $A$.

\subsection{Type checking under cofibrations}\label{sub:tyck-cofib}
We start from an example of type checking a partial element.
\begin{example}\label{ex:parTyck}
Consider $\Psi:=x:\II, y:\II,z:\II$ and the following:
\begin{mathpar}
\inferrule{}{\Gvdash\isType{A}} \and
\inferrule{}{\Gvdash u:\II\to\II\to A}\and
\inferrule{}{\Gvdash v:\II\to\II\to A}
\end{mathpar}
We want to construct the following partial element (recall~\cref{rem:sq-ori}):
\carloCTikZ{\carloXyz \refcube{BBFaces}}
Recall~\cref{fig:parEl}, we may directly translate that into:
\[\LRbbar{\begin{array}{rll}
z&=\rcon&\mapsto u~x~y\\
y&=\lcon&\mapsto v~x~z
\end{array}}\]
Note that they share the same line \fbox{$z=\rcon\land y=\lcon$}
(obtained by taking the conjunction of both cofibrations).
This line is represented as a cofibration, and we can also use it as
a \textit{substitution}, i.e. $[\rcon/z,\lcon/y]$.
Substituting with this line corresponds to the operation of
taking this line from a cube (that has this line).
Since both faces have this line, we may represent the shared line
by substituting either of them:
\begin{align*}
(u~x~y)[\rcon/z,\lcon/y]&\implies u~x~\lcon\\
(v~x~z)[\rcon/z,\lcon/y]&\implies v~x~\rcon
\end{align*}
However, these two lines are actually the same line,
so they have to be definitionally the same, i.e. $u~x~\lcon\equiv v~x~\rcon$.
In this case, we say that the given two faces \textit{agree}.
For a well-defined partial element, every pair of faces should agree.
The typing rule ends up like this:
\begin{mathpar}
\inferrule{\Hint{\Gvdash\isType{A}}\\
\PGvdash u~x~y:A\\
\PGvdash v~x~z:A\\\\
\PGvdash u~x~\lcon\equiv v~x~\rcon:A}{\PGvdash\LRbbar{\begin{array}{rll}
z&=\rcon&\mapsto u~x~y\\
y&=\lcon&\mapsto v~x~z
\end{array}}:\Partial{(z=\rcon\lor y=\lcon)}A}
\end{mathpar}
\end{example}

We generalize~\cref{ex:parTyck} to an arbitrary partial element (let $i\in I$ for some index set $I$):
\[\lrbbar{\overline{\conj_i\mapsto u_i}}\]
We need to find an algorithm that makes sure every pair of faces agree.
For two faces $\conj_i\mapsto u_i$ and $\conj_j\mapsto u_j$,
from~\cref{ex:parTyck} we know that they overlap at $\conj_i\land\conj_j$
(in case they do not overlap, this cofibration is self-contradictory).
The rule becomes something like:
\begin{mathpar}
\inferrule{\Hint{\Gvdash\isType{A}}\\
\forall i\in I.(\Psi\vdash \conj_i)\\
\forall i\in I.(\PGvdash u_i:A)\\\\
\forall i,j\in I.(\PGvdash[\text{overlap-check}])}
{\PGvdash\lrbbar{\overline{\conj_i\mapsto u_i}}
:\Partial{\left(\bigvee\nolimits_{i\in I} \conj_i\right)}A}
\end{mathpar}
In~\cref{ex:parTyck}, the overlap-check step is a conversion check
between a \textit{substituted version} of the two faces $u_i$ and $u_j$.
For convenience, we introduce the typing judgment as in~\cref{fig:conv-cofib}.
\begin{figure}[h!]
\[\PGfvdash\conj u\equiv v:A\]
\caption{Conversion check under a conjunction cofibration}\label{fig:conv-cofib}
\end{figure}

To implement this judgment, we convert $\conj$ into a substitution
(this is possible due to~\cref{prop:unique}),
apply to the three terms on the right-hand side of $\vdash$,
and then apply the normal conversion check under $\Psi;\Gamma$.
The idea in~\cref{fig:conv-cofib} can be extended in the following ways:
\begin{itemize}
\item To do a normal type-check under conjunctions,
e.g. \fbox{$\PGfvdash\conj u:A$}, we apply the substitution to $A$
and run the normal type-check.
\item To check anything that results in a \textit{yes} or \textit{no},
we may check them under a disjunction cofibration $\disj$
by iterating the conjunctions in $\disj$ and run the check under that conjunction.
We return \textit{yes} only when all these checks return \textit{yes},
and return \textit{no} otherwise.
\item To check if two cofibrations $\disj_0,\disj_1$ are equivalent.
We simply check if they satisfy each other. So, $\vdash \disj_0\equiv\disj_1$
is equivalent to $\disj_0\vdash \disj_1\equiv\top$ and $\disj_1\vdash \disj_0\equiv\top$.
\end{itemize}
These ideas give rise to the judgments in~\cref{fig:tyck-w-cofib}.
\begin{figure}[h!]
\begin{align*}
\PGfvdash{\disj&} u\equiv v:A\\
\PGfvdash{\conj&} u:A\\
\PGfvdash{\disj&} \isType{A\equiv B}\\
\PGfvdash{\conj&} \isType{A}\\
\PGfvdash{\disj&} \disj_0\equiv \disj_1
\end{align*}
\caption{Judgments under cofibrations}\label{fig:tyck-w-cofib}
\end{figure}

With all these preparations we can get the typing rule
for partial elements (note that we also add support for $A$ to
reference variables in $\Psi$), as in~\cref{fig:tyck-par}.
\begin{figure}[h!]
\begin{mathpar}
\inferrule{\Psi\vdash \disj\\ \PGvdash\isType{A}}{\PGvdash \isTypeBox{\Partial{\disj}A}}\and
\inferrule{\PGvdash u:A}{\PGvdash \lrbbar u : \Partial\top A}\and
\inferrule{
\forall i\in I.(\Psi\vdash \conj_i)\\
\forall i\in I.(\PGfvdash{\conj_i} u_i:A)\\\\
\forall i,j\in I.(\PGfvdash{\conj_i\land\conj_j} u_i\equiv u_j:A)}
{\PGvdash\lrbbar{\overline{\conj_i\mapsto u_i}}
:\Partial{\left(\bigvee\nolimits_{i\in I} \conj_i\right)}A}
\end{mathpar}
\caption{Typing rule of partial elements}\label{fig:tyck-par}
\end{figure}

\section{Applications of partial elements}

\subsection{Generalized paths}\label{sub:extTy}
In~\cite{CCHM}, the path type is defined using the following rules
(slightly paraphrased for notational consistency):
\begin{mathpar}
\inferrule{\PiGvdash x\isType A\\ \PGvdash a:A[\lcon/x]\\ \PGvdash b:A[\rcon/x]}
{\PGvdash\isTypeBox{\PathTy{\lam x A} a b}}\and
\inferrule{\PiGvdash x u:A}
{\PGvdash \plam x u : \PathTy{\lam x A}{(u[\lcon/x])}{(u[\rcon/x])}}\and
\inferrule{\Hint{\PiGvdash x\isType A}\\\PiGvdash x u:A\\\PGvdash v:\II}{
\PGvdash \papp{(\plam x u)}v \equiv u[v/x]:A[v/x]}\and
\inferrule{\Hint{\PiGvdash x\isType A}\\\PGvdash u:\PathTy{\lam x A} a b}{
\PGvdash \papp u\lcon \equiv a:A[\lcon/x] \\ \text{and} \\ \PGvdash \papp u\rcon \equiv b:A[\rcon/x]}
\end{mathpar}
We may rephrase these rules using the idea of cofibrations and partial elements:
\begin{itemize}
\item The path type \fbox{$\PathTy{\lam x A} a b$}
is similar to a $\Pi$-type \fbox{$(x:\II)\to A$} carrying a partial element.
We denote the whole thing as:
\[\ExtTy{x}{A}{rll}
{x&=\lcon&\mapsto a\\
x&=\rcon&\mapsto b}\]
\item The introduction rule takes a type
\fbox{$\extTy{x}{A}{\overline{\conj_i\mapsto v_i}}$} and checks the following:
\begin{itemize}
\item \fbox{$\PiGvdash x u:A$}, just like checking \fbox{$\lam x u:(x:\II)\to A$}.
\item \fbox{$\forall i. (\Psi,x:\II;\Gamma,\conj_i\vdash u\equiv v_i:A)$},
like saying ``$u$ \textit{matches} every given face''.
\end{itemize}
\item The elimination rule is the same as $\Pi$-types.
\item The computation rules combine the computation rule of $\Pi$-types
and the reduction of partial elements (defined in~\cref{sub:red-cofib}).
\end{itemize}
In particular, we can relax the formation rule to allow multiple intervals
and an arbitrary partial element (but the cofibrations may only use the
provided intervals), and the rest of the rules will still work.
This gives us a better definition of the path type,
which is more convenient to work with.
We may specify only one endpoint (unlike a traditional path,
which always requires two endpoints) of a generalized path as in~\cref{ex:gconcat},
or specify the inner boundaries of squares as in~\cref{ex:gsquare}.

\begin{terminology}
The original idea of generalized paths was introduced in~\cite[\S 2.2 and Fig. 4]{InfCat},
which uses the name ``extension types''. It is also implemented in the proof assistants
\texttt{\textcolor{red}{red}tt}\footnote{\url{https://github.com/RedPRL/redtt}} and
\texttt{\textcolor{blue}{cool}tt}\footnote{\url{https://github.com/RedPRL/cooltt}}.
\end{terminology}

We extend the syntax with generalized paths in~\cref{fig:path}.
\begin{figure}[h!]
\[u,A::= \extTy{\overline{x}}{A}{\overline{\face}}
\mid \plam x u \mid \papp u v
\mid \cdots~\text{(see~\cref{fig:parTy})}\]
\caption{Syntax of generalized paths}\label{fig:path}
\end{figure}

The typing rules are defined in~\cref{fig:tyck-path}.
Note that we make use of reduction on partial elements (\cref{sub:red-cofib}).
\begin{figure}[h!]
\begin{mathpar}
% Formation
\inferrule{\Psi\overline{,x:\II};\Gvdash \isType{A}\\
\forall i.(\overline{x:\II}\vdash\conj_i) \\\\
\PGvdash\lrbbar{\overline{\conj_i\mapsto u_i}}
:\Partial{\left(\bigvee\nolimits_{i} \conj_i\right)}A}
{\PGvdash\isTypeBox{\extTy{\overline{x}}{A}{\overline{\conj_i\mapsto u_i}}}}
\and
% Introduction
\inferrule{\Psi\overline{,x_i:\II};\Gamma\vdash v:A[\overline{x_i/y_i}]\\\\
\forall j. (\Psi\overline{,x_i:\II};\Gamma,\conj_j[\overline{x_i/y_i}]\vdash v\equiv u_j[\overline{x_i/y_i}]:A[\overline{x_i/y_i}])}
{\PGvdash \plam{\overline{x_i}}{v}:\extTy{\overline{y_i}}{A}{\overline{\conj_j\mapsto u_j}}}
\and
% Elimination
\inferrule{\PGvdash \overline {u_i:\II}\\\\
\PGvdash v:\extTy{\overline{x_i}}{A}{\overline\face}}
{\PGvdash \papp v {\overline{u_i}} : A[\overline{u_i/x_i}]}
\and
% Computation
\inferrule{\PGvdash \overline {u_i:\II}\\
\Psi\overline{,x_i:\II};\Gamma,\vdash v:A}
{\PGvdash \papp{(\plam{\overline{x_i}}v)}{\overline{u_i}}\equiv
v[\overline{u_i/x_i}] : A[\overline{u_i/x_i}]}
\and
\inferrule{\PGvdash \overline {u_i:\II}\\
\PGvdash v:\extTy{\overline{x_i}}{A}{\overline\face}\\\\
\PGvdash \lrbbar{\overline\face}[\overline{u_i/x_i}]\equiv \lrbbar{v_0} : A[\overline{u_i/x_i}]}
{\PGvdash \papp v {\overline{u_i}}\equiv v_0 : A[\overline{u_i/x_i}]}
\end{mathpar}
\caption{Typing rules of generalized paths}\label{fig:tyck-path}
\end{figure}

\begin{remark}
With named bindings, the following criterion from~\cref{fig:tyck-path}:
\[\Psi\overline{,x_i:\II};\Gamma,\conj_j[\overline{x_i/y_i}]\vdash v\equiv u_j[\overline{x_i/y_i}]:A[\overline{x_i/y_i}]\]
has an alternatively implementation:
\[\Psi\overline{,y_i:\II};\Gamma,\conj_j\vdash v[\overline{y_i/x_i}]\equiv u_j:A\]
The idea is that we may substitute the term to use the bindings in the type and do the conversion checks,
instead of substituting the type to the use the bindings in the term.
Note that this difference does not exist with index-based bindings.

The alternative implementation is shorter, but the philosophy behind it is very bad:
it is usually preferred to adapt the type during the type checking of a term,
not the other way around. We may want to store a proof of the fact that the term is well-typed,
and it's better if the proof is about the term itself, not a substituted version of it.
During the development of \GuestName, the alternative way was first used and then replaced by the
philosophically more faithful way.
\end{remark}

\begin{example}\label{ex:gconcat}
The path concatenation operation can have a simpler type signature.
With the path type, its type is:
\[\textsf{concat}~(a:A)~(b:A)~(c:A)~(p:\PathTy{\lam \_ A} a b)
~(q:\PathTy{\lam \_ A} b c):\PathTy{\lam \_ A} a c\]
The first three parameters can be inferred from the last two,
but we have to quantify over them anyway.
Using the generalized path type, we may simplify it as:
\begin{align*}
\textsf{concat}~&(p:\extTy{\_}{A}{})~(q:\extTy{\_}{A}{x=\lcon\mapsto p~@~\rcon})\\
&:{\ExtTy{\_}{A}{rll}{x&=\lcon&\mapsto p~@~\lcon\\
x&=\rcon&\mapsto q~@~\rcon}}
\end{align*}
It becomes longer, but the new definition has fewer parameters.
The new definition can be more friendly to program synthesizers
or programming languages that do not support implicit arguments.
\end{example}

\begin{example}\label{ex:gsquare}
Consider \fbox{$p_i:\PathTy{\lam\_ A} \bullet \bullet$} for $i\in\set{0,1,2,3}$
where \fbox{$\bullet:A$} and the following square \fbox{$x:\II,y:\II\vdash u$}:
\carloCTikZ{\carloXy
\carloSqBullets
\fill [pattern color=lightgray,pattern=north west lines] (0,0) rectangle (1,1) ;
\draw[->] (00) -- (01) node [midway,left] {$\papp{p_2}y$} ;
\draw[->] (10) -- (11) node [midway,right] {$\papp{p_3}y$} ;
\draw[->] (00) -- (10) node [midway,above] {$\papp{p_0}x$} ;
\draw[->] (01) -- (11) node [midway,below] {$\papp{p_1}x$} ;
\node (center) at (0.5, 0.5) {$u$} ;}
Traditionally, $u$ has the following type:
\[\PathTy{\lam y{(\PathTy{(\lam\_ A)}{(\papp{p_2}y)}{(\papp{p_3}y)})}}{p_0}{p_1}\]
The generalized version is apparently more straightforward:
\[\ExtTy{x,y}{A}{rll}{
x&=\lcon&\mapsto \papp{p_2}y\\
x&=\rcon&\mapsto \papp{p_3}y\\
y&=\lcon&\mapsto \papp{p_0}x\\
y&=\rcon&\mapsto \papp{p_1}x}\]
\end{example}

\begin{remark}
It would be nice if we can implement automatic coercion from
pi types $\overline{(x:\II)}\to X$ to generalized path types
and vice versa by $\eta$-expansion.
This makes programming with \CTT{} so much easier.
\end{remark}

\subsection{Cubical subtypes}\label{sub:cubsub}
Cofibrations in generalized paths are always \textit{bounded}
in the sense that they cannot reference arbitrary local variables.
We add the \textit{open} version of generalized paths,
called \textit{cubical subtypes}, into our type theory.
The syntax is defined in~\cref{fig:subtype}.
\begin{figure}[h!]
\begin{align*}
	u,A::= &~~ \subTy{A}{\disj}{\overline\face}\\
\mid &~~ \outS\disj u \mid \inS\disj u\\
\mid &~~ \cdots~\text{(see~\cref{fig:path})}
\end{align*}
\caption{Syntax of cubical subtypes}\label{fig:subtype}
\end{figure}

The type \fbox{$\subTy{A}{\disj}{\overline\face}$} has three parameters: the \textit{ambient} type $A$,
a cofibration $\disj$, and a set of faces ${\overline\face}$ such that:
\[\lrbbar{\overline\face}:\Partial{\disj}A\]
Then, if \fbox{$\PGvdash u:A$} and $u$ \textit{matches} all the given faces
(``match'' as in introduction of generalized paths), then:
\[\inS\disj u:\subTy{A}{\disj}{\overline\face}\]
This gives the introduction rule.
The elimination rule is just the obvious inverse of the introduction rule,
denoted \outSLong.
Hence the typing rules as in~\cref{fig:typing-subtype}.

\begin{figure}[h!]
\begin{mathpar}
\inferrule{\Hint{\PGvdash \isType A} \\
\PGvdash \lrbbar{\overline\face}:\Partial{\disj}A}
{\PGvdash\isTypeBox{\subTy{A}{\disj}{\overline\face}}}\and
\inferrule{\forall i.(\PGfvdash{\conj_i} u\equiv v_i)}
{\PGvdash\inS\disj u:\subTy{A}{\disj}{\overline{\conj_i\mapsto v_i}}}\and
\inferrule{\PGvdash u:\subTy{A}{\disj}{\overline\face}}
{\PGvdash\outS\disj u:A}
\end{mathpar}
\caption{Typing rules of cubical subtypes, typing part}\label{fig:typing-subtype}
\end{figure}

Cubical subtypes compute in the following cases:
\begin{itemize}
\item The juxtaposition of elimination and introduction always cancels each other.
By type checking, they should have the same cofibrations.
\item The elimination rule computes according to the partial element.
\end{itemize}
Hence the computation rules as in~\cref{fig:compute-subtype}.
\begin{figure}[h!]
\begin{mathpar}
\inferrule{\PGvdash u:A \\ \PGvdash \disj \\ \PGvdash \disj}
{\PGvdash \outS{\disj}{\inS{\disj} u}\equiv u:A} \and
\inferrule{\PGvdash u:\subTy{A}{\disj}{\overline\face}}
{\PGvdash \inS{\disj}{\outS{\disj} u}\equiv u:\subTy{A}{\disj}{\overline\face}}\and
\inferrule{\PGvdash u:\subTy{A}{\disj}{\overline{\conj_i\mapsto v_i}}}
{\forall i.(\PGfvdash{\conj_i} \outS{\disj}u\equiv v_i:A)}
\end{mathpar}
\caption{Typing rules of cubical subtypes, computation part}\label{fig:compute-subtype}
\end{figure}

The set of instances of this cubical subtype is similar\footnote
{It would be nice to say \textit{canonically isomorphic to} here, but the word
\textit{canonical} is being criticized on social media recently,
so it's avoided.} to a subset of the instances of $A$
such that these instances match the given faces.

\begin{remark}
Ideally, we would define cubical subtypes to be \textit{subtypes}
in the sense of subtyping, instead of being explicitly coerced.
However, subtyping in dependent type theories is extremely
sophisticated and may have bad consequences.
\end{remark}

Cubical subtypes, partial elements, and generalized paths are all
based on the idea of cofibrations. Here's a brief table for comparison:

\begin{center}
\begin{tabular}{|l|c|c|c|}
 \hline
 & \textsf{Sub} & \textsf{Partial} & \textsf{Ext} \\ \hline
 Have faces in type? & \checkmark & $\times$ & \checkmark \\ \hline
 Have faces in terms? & $\times$ & \checkmark & $\times$ \\ \hline
 Scope of cofibrations & Open & Open & Bound \\ \hline
%  Number of faces & $1$ & $\geq 1$ & $\geq 0$ \\ \hline
%  Support composition & $\times$ & $\times$ & \checkmark \\ \hline
%  Support coercion & $\times$ & $\times$ & \checkmark \\ \hline
\end{tabular}
\end{center}

Cubical subtypes will be useful later in~\cref{sub:transp}.

\section{Coercion rules}
This section introduces the notion of \textit{coercion rules} and
motivates the De Morgan structure on the interval type.

\subsection{Minimum and maximum squares}\label{sub:minmax}
Consider a path \fbox{$p:\PathTy{\lam\_ A} a b$},
it makes geometric sense to have some path between $p$ and the identity path
on either $a$ or $b$, because they are all connected together.
In \CTT{} we introduce the operators $\lor$ and $\land$,
which are binary operators that return the maximum and minimum of the operands,
respectively. In~\cite[\S 3]{CCHM}, this construction is described as a
``free distributive lattice generated by symbols''.

The rules of $\lor$ and $\land$ are very simple, but listed in~\cref{fig:minmax-rules}.
\begin{figure}[h!]
\begin{mathpar}
\inferrule{\PGvdash u:\II \\\\ \PGvdash v:\II}{\PGvdash u\land v:\II}\and
\inferrule{\PGvdash u:\II}{\PGvdash u\land \lcon\equiv \lcon:\II}\and
\inferrule{\PGvdash u:\II}{\PGvdash u\land \rcon\equiv u:\II}\and
\inferrule{\PGvdash u:\II \\\\ \PGvdash v:\II}{\PGvdash u\lor v:\II}\and
\inferrule{\PGvdash u:\II}{\PGvdash u\lor \lcon\equiv u:\II}\and
\inferrule{\PGvdash u:\II}{\PGvdash u\lor \rcon\equiv \rcon:\II}\and
\inferrule{\PGvdash u:\II}{\PGvdash u\lor u\equiv u:\II}\and
\inferrule{\PGvdash u:\II}{\PGvdash u\land u\equiv u:\II}
\end{mathpar}
\caption{Typing and computation rules of $\land$ and $\lor$}
\label{fig:minmax-rules}
\end{figure}

With these operators we may construct the following squares:
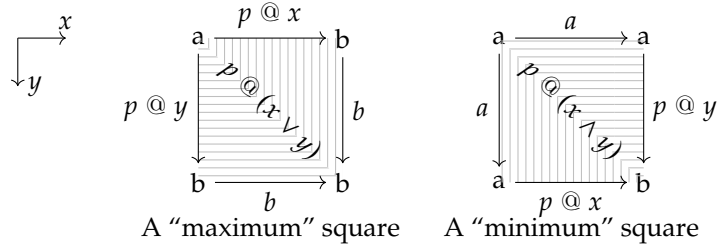
\newcommand{\minmaxPoints}[3]{
\node (00) at (0, 0) {a};
\node (11) at (1.2, 1.2) {b};
\node (10) at (1.2, 0) {#1};
\node (01) at (0, 1.2) {#1};
\fill[pattern color=lightgray,pattern=horizontal lines]
  (00.center) -- (11.center) -- (#2.center) -- cycle ;
\fill[pattern color=lightgray,pattern=vertical lines]
  (00.center) -- (11.center) -- (#3.center) -- cycle ;
}
\newcommand{\minmaxCenter}[2]{
\node[rotate=-45] (center) at (0.6, 0.6) {$\papp p{(#1)}$} ;
\node (text) at (0.6, 1.6) {A ``#2'' square} ;
}
\carloCTikZ{\carloXy
\minmaxPoints{b}{01}{10}
\draw[->] (00) -- (01) node [midway,left] {$\papp p y$} ;
\draw[->] (10) -- (11) node [midway,right] {$b$} ;
\draw[->] (00) -- (10) node [midway,above] {$\papp p x$} ;
\draw[->] (01) -- (11) node [midway,below] {$b$} ;
\minmaxCenter{x\lor y}{maximum}
\shiftTikZ{2.5, 0}{
\minmaxPoints{a}{10}{01}
\draw[->] (00) -- (01) node [midway,left] {$a$} ;
\draw[->] (10) -- (11) node [midway,right] {$\papp p y$} ;
\draw[->] (00) -- (10) node [midway,above] {$a$} ;
\draw[->] (01) -- (11) node [midway,below] {$\papp p x$} ;
\minmaxCenter{x\land y}{minimum}
}}

We also introduce the $\neg$ operator, which is a unary operator that returns
the point of symmetry on $\II$, hence the De Morgan laws are applicable to $\neg$,
$\lor$, and $\land$. The rules are like:
\begin{align*}
	\neg(u\land v)&\equiv \neg u \lor \neg v\\
	\neg(u\lor v)&\equiv \neg u \land \neg v\\
\end{align*}
Some other obvious rules on $\neg$ include $\neg(\neg u)\equiv u$,
$\neg\lcon\equiv\rcon$, and $\neg\rcon\equiv\lcon$.

\begin{exercise}
Think about the squares we can construct with a path
$p:\PathTy{\lam\_A}a b$ and the De Morgan structures on $\II$,
like in~\cref{rem:sq-ori}.
\end{exercise}

\begin{exercise}\label{ex:norm-i}
Show that terms of type $\II$ (generated by variables, $\neg$, $\land$, and $\lor$)
strongly normalizes to unique normal forms.
This implies that expressions of type $\II$ in open contexts has a
decidable conversion check algorithm.
\end{exercise}

Note that there are established facts that support~\cref{ex:norm-i},
so it is not recommended to do it in a brute-force way.

\begin{lem}\label{lem:isomorphism}
There exists an isomorphism between expressions of type $\II$ in open contexts
and cofibrations in the same context.
\end{lem}
\begin{proof}
By constructing the following bijective function.
\[\begin{array}{rl}
\phi(x)&:=x=\rcon\\
\phi(\neg x)&:=x=\lcon\\
\phi(x\land y)&:=\phi(x)\land\phi(y)\\
\phi(x\lor y)&:=\phi(x)\lor\phi(y)
\end{array}\]
It is left to the reader to construct its inverse.
\end{proof}

\subsection{Freezing}\label{sub:freeze}
\begin{defn}[Freeze]\label{defn:freeze}
A type-line $A:\II\to\UU$ \textit{freezes} on a cofibration $\disj$ if
it is \textit{constant} under $\disj$. In terms of typing judgment:
\[\Psi,x:\II;\Gamma,\disj\vdash \isTypeBox{A~\lcon\equiv A~x}\]
The judgment is from~\cite[\S 2.1]{HCompPDF}.
\end{defn}
Here is an illustration of~\cref{defn:freeze} inspired from an online cubical tutorial
by Favonia\footnote{\url{https://youtu.be/uE6g4Oyom68?t=710}}.
Consider the type-line $A:\II\to\UU$, how can it be frozen under a cofibration?
\begin{enumerate}
\item If $A$ is already constant,
then it is frozen under all cofibrations, which is not interesting.
\item Consider a non-constant $A$ which is constant under $\disj$.
For convenience, let $A:=\plam x{A_0}$. Apparently,
there are subexpressions in $A_0$ involving the interval variable $x$.
\item Recall~\cref{sub:tyck-cofib}, this means under the interval substitution
that corresponds to $\disj$, these subexpressions will reduce
and the result does not refer to $x$.
\item So, these subexpressions have to refer to interval variables other than $x$
which are substituted away under $\disj$!
\end{enumerate}
Hence the following example.
\begin{example}
Suppose $A_0$ also refers to $y:\II$. This makes $A$ a square:
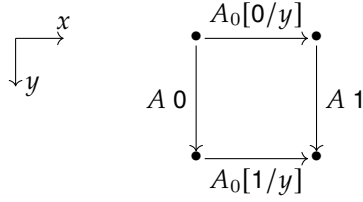
\carloCTikZ{\carloXy
\carloSqBullets
\draw[->] (00) -- (01) node [midway,left] {$A~\lcon$} ;
\draw[->] (10) -- (11) node [midway,right] {$A~\rcon$} ;
\draw[->] (00) -- (10) node [midway,above] {$A_0[\lcon/y]$} ;
\draw[->] (01) -- (11) node [midway,below] {$A_0[\rcon/y]$} ;}
Since $\disj$ substitutes $y$ away, either $y=\lcon\in\disj$ or $y=\rcon\in\disj$.
We may choose either for demonstration purpose, say,
$y=\lcon$. We may also forget about the other conditions,
as they are unrelated to the square. Then, since $A$ \textit{freezes} under $y=\lcon$,
the corresponding face must be constant, resulting in a triangle:
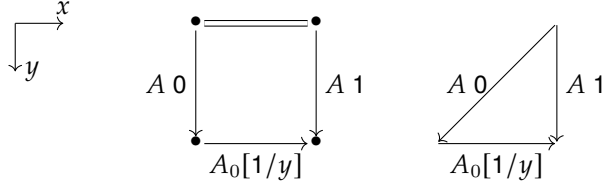
\carloCTikZ{\carloXy
\carloSqBullets
\draw[->] (00) -- (01) node [midway,left] {$A~\lcon$} ;
\draw[->] (10) -- (11) node [midway,right] {$A~\rcon$} ;
\draw[equals arrow] (00) -- (10) node [midway,above] {} ;
\draw[->] (01) -- (11) node [midway,below] {$A_0[\rcon/y]$} ;
\shiftTikZ{2, 0}{
\carloSqVertices{~~}
\draw[->] (10) -- (01) node [midway,left] {$A~\lcon$} ;
\draw[->] (10) -- (11) node [midway,right] {$A~\rcon$} ;
\draw[->] (01) -- (11) node [midway,below] {$A_0[\rcon/y]$} ;
}}
In short, $A$ freezes under $\disj$ if the faces on $A$
specified by $\disj$ are constant.
\end{example}
\begin{remark}
Note that the judgment in~\cref{defn:freeze} is not equivalent to:
\[\PGfvdash\disj \isTypeBox{A~\lcon\equiv A~\rcon}\]
Because a nontrivial loop (hence the two sides are the same,
but it's distinguishable from the identity path, hence not constant)
satisfies the latter condition.
\end{remark}
Here are some simple facts that might be helpful for building up
intuition about freezing:
\begin{lem}
Any type line freezes on the absurd cofibration.
\end{lem}
\begin{lem}
If a type line freezes on the truth cofibration,
then it is constant.
\end{lem}

\subsection{The coercion operator}\label{sub:transp}
In \CTT, for every type-line \fbox{$A:\II\to\UU$},
we add a function $\coeLong:A~\lcon\to A~\rcon$ to the type theory.
Since a type-line expresses the equivalence between two types (the two sides
of the line), the instances of these two types can be safely coerced.

% \begin{remark}
% The fact that all functions from $\II$ induce an equivalence is
% related to the property of $\II$ known as \textit{tininess}~\cite{AmazingRightAdjoint}.
% \end{remark}

We expect coercion to satisfy the following properties:
\begin{itemize}
\item It is an isomorphism -- so an inverse $\coeLong_A^{-1}:A~\rcon\to A~\lcon$ must exist.
One may easily guess that this is induced by the
$\coeLong$ function on the type-line $\plam x A~(\neg x)$.
\item There exists a cofibration $\disj$ that $A$ freezes (\cref{defn:freeze}) on it.
Furthermore, we want that for every $u:A~\lcon$:
\[\disj\vdash \coeLong(u)\equiv u:A~\lcon\]
\item It computes on type formers, e.g. for the type line $\lam x{A~x\to B~x}$, we have:
\[\coeLong:(A~\lcon\to B~\lcon)\to(A~\rcon\to B~\rcon)\]
And we expect this function to behave like:
\begin{align*}
&\text{take}~f:A~\lcon\to B~\lcon~\text{and}~a:A~\rcon&&\text{from input}\\
\implies &\coeLong^{-1}(a):A~\lcon&&\text{by type line}~\lam x{A~x}\\
\implies &f(\coeLong^{-1}(a)):B~\lcon&&\text{by function application}\\
\implies &\coeLong(f(\coeLong^{-1}(a))):B~\rcon&&\text{by type line}~\lam x{B~x}\\
\end{align*}
Hence the well-typed result: an instance of type $B~\rcon$.
This means the function does not obviously break canonicity.
\end{itemize}
The behavior of $\coeLong$ is considered an extra set of \textit{rules}
called \textit{coercion rules} on each type former,
just like introduction rules, elimination rules,
and computation rules.

\begin{terminology}
In~\cite{CCHM,CHM,HCompPDF}, coercion is called \textit{transport}
or \textit{generalized transport}.
We use the name \textit{coercion} (inspired from the Cartesian cubical
type theory literature, including~\cite{ABCFHL,CoolCCTT}) because it fits
better with the traditional functional programming literature.
\end{terminology}

To achieve all the properties we have asked for,
we add the following ``function'' to \CTT (\cref{fig:syntax-transp}):

\begin{figure}[h!]
\[u,A::= \coe{A}{\disj} \mid \cdots~\text{(see~\cref{fig:subtype})}\]
\caption{Syntax of coercion}
\label{fig:syntax-transp}
\end{figure}

It is actually not the normal sense of function,
because the function body can't be expressed internally in the type theory.
It is more like an operator whose behavior is induced by the type $A$.

The implementation is basically a rewrite of~\cite[\S 2.1, \S 3]{HCompPDF},
as in~\cref{fig:typing-transp}.

\begin{figure}[h!]
\begin{mathpar}
\inferrule{\PGvdash A:\II\to\UU \\
\Psi\vdash\overline{\conj_i} \\
\Psi,x:\II;\Gamma,\overline{\conj_i}\vdash \isTypeBox{A~x\equiv A~\lcon}}{
\coe{A}{\overline{\conj_i}}:(u:A~\lcon) \to
 \subTy{(A~\rcon)}{\overline{\conj_i}}{\overline{\conj_i\mapsto u}}}
\end{mathpar}
\caption{Typing of the coercion rule}
\label{fig:typing-transp}
\end{figure}

The type signature is a bit complicated because of the cubical subtype
used to express the redution of $\coeLong$. We may simplify by separating the
computation behavior from the typing, as in~\cref{fig:typing-transp-2}.
This approach does not have the cubical subtype boilerplate, which leads to cleaner code.

\begin{figure}[h!]
\begin{mathpar}
\inferrule{\PGvdash A:\II\to\UU \\
\Psi\vdash\disj \\
\Psi,x:\II;\Gamma,\overline{\conj_i}\vdash\isTypeBox{A~x\equiv A~\lcon}}{
\PGvdash\coe{A}{\disj}:A~\lcon \to A~\rcon}\and
\inferrule{\PGvdash A:\II\to\UU \\ \Psi\vdash\disj}{
\PGfvdash\disj\fbox{$\coe{A}{\disj}\equiv \lam x x$}:
A~\lcon \to A~\lcon\footnotemark}
\end{mathpar}
\caption{Alternative way to type the coercion rule}
\label{fig:typing-transp-2}
\end{figure}
\footnotetext{Note that the conversion rule is well-typed because under $\disj$, $A~\lcon\equiv A~\rcon$}

\begin{remark}
The work~\cite{CubicalAgda} has implemented it as a built-in function
with the following type signature:
\[\coeLong:(A:\II\to\mathbb{U})~(\psi:\II)~(A~\lcon)\to(A~\rcon)\]
However, invocations to this function is valid only when $A$ is constant under $\psi$,
and this criterion is not expressed in the type signature.
It is said to be a design flaw\footnote{\url{https://github.com/agda/agda/issues/3509\#issuecomment-456201777}}
in the Agda bug tracker as well as in~\cite[Cool Remark 2.2]{CoolCCTT}.
The latter will be discussed in~\cref{rem:lame}.
\end{remark}

\begin{lem}[Fill]\label{lem:transpFill}
Given a well-typed term: \[\coe A{\disj}{(u)}\]
There exists the following path: \[p:\ExtTy x {A~x}{ccl}{
	x&=\rcon&\mapsto \coe A{\disj}{(u)} \\
	x&=\lcon&\mapsto u}\]
In more traditional notations:
\[p:\PathTy{A} u {(\coe A{\disj}{(u)})}\]
Diagrammatically:
\[\begin{tikzcd}
	u && {\coe A{\disj}{(u)}}
	\arrow["p"', from=1-1, to=1-3]
\end{tikzcd}\]
\end{lem}
\begin{proof}
Let
\[p:=\plam x{\coe{\lam y{A~(x\land y)}}
 {\disj\lor x=\lcon}(u)}\]
It is left to the reader to check the well-typedness of these constructions
(see discussion in~\cite{HCompPDF} which is extremely brief but still useful).
In some sense, this construction motivates the De Morgan structure on $\II$.
\end{proof}

\begin{terminology}[Fill]
The path $p$ in~\cref{lem:transpFill} is called the \textit{filler} of $\coe A{\disj}$.
In general, for a function $f:A~\lcon\to A~\rcon$ where $A$ is just any type line,
the path $p:u=f(u)$ is called the \textit{filler} of $f$.
We will see the construction of another filler like this later in~\cref{sec:hcomp}.

More generally, given \textit{any} partial configuration,
a cube matching it is called a \textit{filler} (of the partial configuration).
\end{terminology}

The coercion rules for type formers are omitted as they are
very long and are already written in~\cite[\S 3]{HCompPDF}. By comparing~\cref{lem:transpFill}
and~\cite[\S 2.1]{HCompPDF}, one should be able to translate the rest.
There is one exception though -- the coercion rule for the generalized path type
is slightly different. To define the rule, we need another Kan operation.

\section{Composition}\label{sec:hcomp}
There is another primitive construction which is needed to define the
coercion rules for the (generalized) path type:
the \textit{composition} operation, which takes a configuration
of an $n$-cube \textit{without} the top face (also known as an \textit{open box}),
and constructs the missing top face. In other words, composition asserts that any
open box are \textit{closed} (i.e.~the top face is, in fact, filled).

\subsection{Intuition behind composition}
The intuition behind the composition operation is like ``any cube
that \textit{makes cubical sense} has a filling'', which gives \CTT{} the terms
that we desire but cannot construct using other term formers.
The most notable motivating example is path concatenation (\cref{ex:concat-sym}).

There are two versions of composition: \textit{homogeneous composition} (hereafter as \hcompLong) and
\textit{heterogeneous composition}. The former requires that all the given faces
are homogeneous (\cref{defn:homo-path}), while the latter does not have this restriction.
Heterogeneous composition can be implemented in terms of homogeneous composition \textit{and} \coeLong.
This tutorial will work with primitive homogeneous composition~\cref{sub:homo-hcomp},
and define heterogeneous composition as a corollary~\cref{sub:heter-hcomp}.

\begin{terminology}
When referring to particular faces of a cube, we call the front, back, left, and right faces \textit{walls}.
Walls are parts of the (optional) input data of composition.
The top and the bottom faces are either a (mandatory) input or an output of composition,
depending on the authors' preference.
We use the bottom face as the input and call it the \textit{floor}.
In case of higher dimensions, we think of cubes as a term \fbox{$p:\II^n \to X$}
with $p~\lcon$ called the floor and $p~\rcon$ the \textit{top face} or the \textit{ceiling}.
All the other faces are called the walls.

In~\cite{CoolCCTT}, the top face is the input and is called the \textit{cap}
(and the walls are called the \textit{tubes}).
\end{terminology}

\begin{history}
In~\cite{CCHM}, heterogeneous composition \texttt{comp} is the only primitive Kan operation,
with \coeLong{} defined as a special case where the given configuration has ``no walls''.
For higher inductive types convenience, \cite{HCompPDF,CHM} decomposed \texttt{comp} into a
homogeneous version \hcompLong{} (which could possibly be a canonical term) and a primitive \coeLong.
\end{history}

In order to derive a syntax of homogeneous composition, we start from an example.
Homogeneous composition behaves like a function
that takes an ``open box'' of type $\Partial \disj A$ and returns the missing top face,
which is of type $A$. Below is a demonstration with upside-down cubes
(because we want the floors to be more visible):
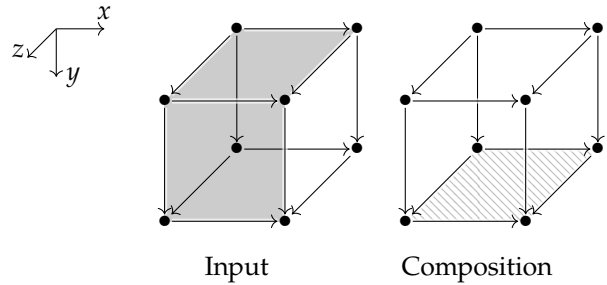
\carloCTikZ{
\carloXyz
\refcube{Partial}
\node (text) at (0, 2) {Input} ;
\shiftTikZ{2,0}{\refcube{Comped}
\node (text) at (0, 2) {Composition} ;}
}
Note that \hcompLong{} works for all $n$-cubes, not just 3-dimensional ones.

\subsection{Homogeneous composition}\label{sub:homo-hcomp}
Given a partial element, it is quite difficult to figure out
which face is the missing one that we would like to compose.
However, it is also very important to make that obvious.
% We would like to make which face is the missing one obvious.
Hence, we need a careful organization of the parameters.

The practice in \CTT{} is to use the type \fbox{$\II \to \Partial \disj A$} to represent the walls,
and ask separately for the floor. This is because the floor is mandatory
while the walls are optional\footnote{This is a questionable design,
speaking of evaluation performance.
However, there isn't any known easy workarounds.
See the ICMS talk, which proposed a ``7-levels type theory'' solution:
\url{https://cse.umn.edu/cs/feature-stories/nullable-compositions-talk}}.
where the parameter is the ``direction of composition'',
which is used to orient the walls (using the idea in~\cref{rem:sq-ori}).
This decomposition requires an additional check that makes sure
the walls and the floor \textit{agree} on the intersection.
For example, the above composition problem
is decomposed as (pictures are still upside-down):
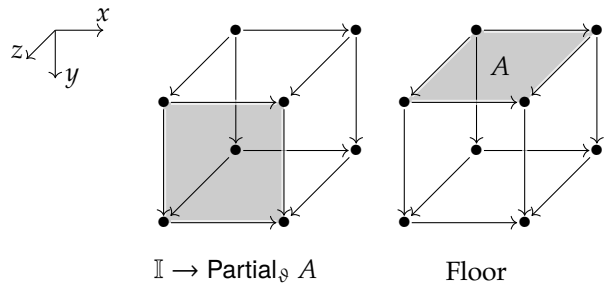
\carloCTikZ{
\carloXyz
\refcube{PartialSide}
\node (text) at (0, 2) {$\II \to \Partial \disj A$} ;
\shiftTikZ{2,0}{\refcube{PartialBot}
\node (text) at (0, 2) {Floor} ;}
}
Note that $\disj$ in this case is $(z=\rcon)$,
and the corresponding terms should look like the following,
where \fbox{$y:\II;z= \rcon \vdash u:A$}:
\[
	\lam y \lrbbar{z=\rcon \mapsto u} :
	\II \to \Partial{z=\rcon} A
\]
Then, \hcompLong{} should give us the following,
where $u[\rcon/y]$ is just the top face of the wall $u$
(it is left to the reader to figure out why the floor
must be $u[\lcon/y]$):
\[
\hcomp{A}{\lam y \lrbbar{z=\rcon \mapsto u}}{u[\lcon/y]}
: \subTy{A}{z=\rcon}{z=\rcon \mapsto u[\rcon/y]}
\]
\carloCTikZ{
\carloXyz
\refcube{URconX}
}
Note that for convenience, we have omitted the cofibration parameter
of \hcompLong{} because it is already given in another argument.
In practice, it might be needed explicitly to assign a type to \hcompLong.
If \hcompLong{} is ever going to have a type, it should be like the following,
where $\phi$ converts interval to the corresponding cofibration (\cref{lem:isomorphism}):
\begin{align*}
	\hcompLong{} :{}& (A : \UU)~(x:\II)\\
	&(w:\II \to \Partial{\phi(x)} A)\\
	&(u:\subTyImpl{A}{\phi(x)}{(w~\lcon)})\\
	\to{}& \subTyImpl{A}{\phi(x)}{(w~\rcon)}
\end{align*}

\begin{remark}[Lame]\label{rem:lame}
The above type signature is a refined version of the one implemented in Agda~\cite{CubicalAgda},
where instead of asking for \fbox{$u:\subTyImpl{A}{\phi(x)}{(w~\lcon)}$}, they simply ask for
\fbox{$u:A$} and inserts a cubical subtype constraint.
In~\cite[Cool Remark 2.2]{CoolCCTT}, the authors claim that the Agda design of \hcompLong{}
is ``unnecessary and inelegant in general'' because the check is ad-hoc and there is obviously
a simpler type signature where $u$ is merged into the walls:
\begin{align*}
\hcompLong{} :{}& (A : \UU)~(x:\II) & \hcompLong{} :{}& (A : \UU)~(x:\II) \\
&(w:\diffA\II \to \Partial{\diffA{\phi(x)}}{A}) & &(w:\diffB{(i:\II)} \to \Partial{\diffB{\phi(x)\lor(i=\lcon)}}{A})\\
&\diffA{(u:\subTyImpl{\diffA A}{\phi(x)}{(w~\lcon)})} & & \\
\to{}& \subTyImpl{A}{\phi(x)}{(w~\rcon)} & \to{}& \subTyImpl{A}{\phi(x)}{(w~\rcon)}
\end{align*}
However, there is a trick that could possibly simplify the reduction rules of \hcompLong{} on
inductive types that would become difficult with the above signature. So we will stick to it
in this tutorial. The readers should be able to translate them back and forth easily.
\end{remark}
The formal syntax and typing of \hcompLong{} is straightforward
(\cref{fig:syntax-hcomp,fig:typing-hcomp}).

\begin{figure}[h!]
\[u,A::= \hcomp A u v \mid \cdots~\text{(see~\cref{fig:syntax-transp})}\]
\caption{Syntax of homogeneous composition}
\label{fig:syntax-hcomp}
\end{figure}

\begin{figure}[h!]
\begin{mathpar}
\inferrule{\PGvdash A:\UU \\
\PGvdash{\disj} \\\\
\PGvdash u : A\\
\PGvdash w : \II \to \Partial{\disj} A\\
\PGfvdash\disj \lrbbar u\equiv w~\lcon : A}{
\hcomp A w u:\subTyImpl{A}{\disj}{(u~\rcon)}}
\end{mathpar}
\caption{Typing of the homogeneous composition}
\label{fig:typing-hcomp}
\end{figure}

The reduction of \hcompLong{} is very simple:
\begin{enumerate}
\item In case walls are constant, it reduces to the walls. In other words,
\[\hcomp{A}{\lrbbar{w}}u\mapsto w\]
\item For types with an introduction rule, it is structural over the rules
(see rules in~\cite{HCompPDF}).
\item For higher inductive types, it is canonical.
\item For universes, see~\cref{sec:read}.
\end{enumerate}
This is the last primitive operator we would like to introduce in this tutorial.
It remains to define the coercion rule of the generalized path type!

\subsection{Heterogeneous composition}\label{sub:heter-hcomp}
Like in~\cite{HCompPDF}, we need some extra operations.
First, we need heterogeneous composition, which is a \textit{dependent} version of \hcompLong.
Here's a comparison on the imaginary type signatures:
% A perfect application of \begin{align*}!
\begin{align*}
\hcompLong{} :{}& (A : \diffA\UU)~(x:\II) & \textsf{comp} :{}& (A : \diffB{\II\to\UU})~(x:\II) \\
&(w:\diffA\II \to \Partial{\phi(x)}{\diffA A}) & &(w:\diffB{(i:\II)} \to \Partial{\phi(x)}{\diffB{(A~i)}})\\
&(u:\subTyImpl{\diffA A}{\phi(x)}{(w~\lcon)}) & &(u:\subTyImpl{\diffB{(A~\lcon)}}{\phi(x)}{(w~\lcon)}) \\
\to{}& \subTyImpl{\diffA A}{\phi(x)}{(w~\rcon)} & \to{}& \subTyImpl{\diffB{(A~\rcon)}}{\phi(x)}{(w~\rcon)}
\end{align*}
The implementation strategy is very crude: we first use coercion to \textit{forward} the walls
and the floor to \fbox{$A~\rcon$}, and then do \hcompLong{} on the constant type \fbox{$A~\rcon$}.
This leads to the following definition.

\begin{defn}\label{defn:comp}
The heterogeneous composition is defined as:
\begin{mathpar}
\inferrule{\PGvdash A:\II\to\UU \\
\PGvdash r : \II}
{\textsf{forward}_A^r := \coe{\lam x {A~(x\lor r)}}{r=\rcon}
: A~r \to A~\rcon} \and
\inferrule{\PGvdash A:\II\to\UU \\
\PGvdash u: A~\lcon\\\\
\PGvdash w : (x:\II) \to \Partial{\disj} {(A~x)}\\
\PGfvdash\disj \lrbbar u\equiv w~\lcon : A~\lcon}
{\textsf{comp}_A(w, u):=
\hcomp{A~\rcon}
{\lam x {\textsf{forward}_{\Partial\disj {(A~x)}}^x(w~x)}}
{\textsf{forward}_A^\lcon(u)}
:{A~\rcon}}
\end{mathpar}
% The idea is that we \textit{forward} the heterogeneous composition problem
% (where the partial element has type $A:\II\to\UU$) to a homogeneous one of type $A~\rcon$,
% and the forward operation is implemented as ``coercion to the right''.
\end{defn}

The desired property of \textsf{forward} is the following definitional equality:
\[\textsf{forward}_A^\rcon(u)\equiv u\]

Note that in~\cref{defn:comp}, we have assumed \textsf{forward} to work for partial elements,
which indirectly assumes \coeLong{} on partial elements, which does not exist.
But it really is just structurally applying \coeLong{} to the faces,
and writing the really formal definition would make it less readable.

% End of the story!
Then, we define the coercion rule of 1-dimensional generalized path types:
\[
\coe{\lam y{\extTy{x}{A}{\overline{\conj_i\mapsto u_i}}}}{\bigvee\nolimits_j \conj_j}(p)
:=
\plam{x}{\textsf{comp}_{\lam x A}(\papp p x, \lam y{\lrbbar{
	\overline{\conj_j \mapsto \papp p x},
	\overline{\conj_i \mapsto u_i}}})}
\]
The idea behind the definition, which should be helpful for readers to generalize it
to higher dimensions, lies in the partial element being constructed:
\begin{enumerate}
\item The coercion needs to be frozen on $\bigvee_j \conj_j$, so we add these faces
	to the walls and let them reduce to the floor $\papp p x$.
\item The faces of the path need to be preserved, so we add them to the walls.
\end{enumerate}
That's it! To generalize to higher dimensions, we need to answer the following questions,
which are left to the readers:
\begin{enumerate}
\item What is the direction of composition?
\item What is the input and output type and in what scope are they defined?
\item How to generalize forwarding to higher dimensions?
\item Do we need to generalized homogeneous composition to higher dimensions?
  The answer is no, but why?
\item How to generalize heterogeneous composition to higher dimensions?
  For sure we will be working with $A:\II^n\to \UU$, but how does the type of $w$ change?
\end{enumerate}
Make sure to validate your definition by specializing to 1 or 2-dimensional cases
and check the results.

\begin{terminology}
A type former is said to be \textit{fibrant} if it supports Kan operations
(i.e. have coercion rules and supports homogeneous composition).

Non-fibrant types are sometimes known as \textit{pretypes}.
A type theory that combines pretypes and fibrant types is known as
\textit{2-level type theory}~\cite{2LTT}.

We tend to work with a strict (exact) equality type on pretypes,
while for fibrant types we use path equality.
\end{terminology}
\begin{example}
Fibrant type formers: $\Pi$-types, $\Sigma$-types, $\UU$,
generalized path types, (higher) inductive types.

Non-fibrant type formers\footnote{In the type theory of the Arend language,
the interval type \textit{is} fibrant, but they do not have the notion of
homogeneous composition or coercion rules. Instead,
they have a coercion operator with regularity and the $\beta$-law of univalence.}:
$\II$, $\Partial\disj A$, $\subTyImpl A \disj u$.
\end{example}

\section{Suggestions for further reading}\label{sec:read}
There are several constructs in \CTT{} that are not covered in this tutorial:
\begin{description}
\item[Cartesian cubical type theory] is a variant of \CTT{} discussed in~\cite{ABCFHL},
  where the De Morgan structure is removed but ``diagonal cofibration'' (like $i\equiv j$,
	where $i$ and $j$ are both variables) is added. The diagonal cofibration implies
	an extensional equality on $\II$.
\item[Glue type] is a type former (with an introduction and elimination rule)
  that is similar to \hcompLong{} on the universe, but with the walls replaced
	with equivalences (functions with contractible fibers).
	In fact, \hcompLong{} on universes is reduced to Glue with trivial equivalences.
	Glue type can be used to prove univalence, and is discussed in~\cite{CCHM,ABCFHL,HCompPDF}.
\item[V type] is a special case of Glue type that is sufficient to prove univalence,
  but cannot handle \hcompLong{} on universes, hence a \CTT{} with V type needs
	to handle \hcompLong{} on universes.
	V type is discussed in~\cite{CHTT3,CoolCCTT}.
\item[Cubical computational type theory] is a variant of \CTT{} built in a computational
  (extensional) type theory, discussed in~\cite{CHTT1,CHTT2,CHTT3,CHTT4,CCCTT}.
	It features an exact equality type with equality reflection.
	Jon Sterling mentioned in a talk%
	\footnote{Slides: \url{https://www.jonmsterling.com/slides/sterling:2022:wits.pdf}}
	that this setting is not going to work well
	due to the lack of automated $\eta$-equality support.
\item[Higher inductive types] are a way to define inductive types generated
  by both normal constructors and path constructors (i.e. that returns
	a cubical subtype of the ambient type).
  They are introduced in~\cite{CHM} with a general scheme
	(attach a partial element on a constructor).
	Higher inductive types in cubical computational type theory are discussed in~\cite{HIT-CCTT}.
	Pattern matching on higher inductive types is discussed in~\cite{CubicalAgda}.
\item[Guardedness] in cubical type theory is discussed in~\cite{GCTT,GHIT}. It is useful to
	integrate coinductive types into cubical type theory.
\item[Regularity] is the $\beta$-law of Martin-L\"of's identity type. It is difficult to
  support this rule in \CTT, see~\cite[\S 3.4]{CAThesis}. This seems to be the \textit{only}
	remaining problem of interpreting homotopy type theory using \CTT.
	Note that we have already proved this rule propositionally (\cref{lem:transpFill}),
	but Martin-L\"of type theory and homotopy type theory assumes the judgmental version of this rule.
\item[Universe of partial types] we have not discussed in what universe the partial types belong to.
  They are pretypes, which is discussed in~\cite{2LTT}. This is also how Cubical Agda implements them after
  their paper is published.
\item[Universe of $\II$] neither did we discuss in what universe does $\II$ belong to.
  They are also pretypes, but are quite different pretypes. See Agda issue tracker
	\href{https://github.com/agda/agda/pull/5439}{\#5439}\footnote{\url{https://github.com/agda/agda/pull/5439}}.
\item[Propositional resizing] is an axiom about universes of propositions.
  It corresponds to the ``completeness'' of the universe: for every type $A$
	that we can show it's a proposition,
	there is supposed to be $[A] : \textsf{Prop}$ (please don't take the notation seriously)
	that is logically equivalent to $A$.
	It can drastically increase proof-theoretic strength of \CTT{} when the universe of proposition
	is impredicative.
\end{description}

The source code of the \GuestName{} project is available at
\url{https://github.com/ice1000/guest0x0}. The development is already integrated into
the Aya Prover\footnote{\url{https://www.aya-prover.org}} which is a more mature system.

\subsection{Acknowledgment}
The author is grateful for Jonathan Sterling, Chang Wen, and Kiva Oyama
for reviewing the early versions of this tutorial and discussing ideas about \CTT{}.
Daniel Gratzer corrected a mistake regarding subtype computation
during the HoTT '23 conference.

\printbibliography
\end{document}